\documentclass[aps,prl,twocolumn]{revtex4-1}

\usepackage{graphicx}
\usepackage{color}
\usepackage{amsmath}
\usepackage{amssymb}
\begin{document}
\title{Statistical learnability of nuclear masses}%

\author{A. Idini}
\affiliation{Division of Mathematical Physics, Department of Physics, LTH, Lund University, Post Office Box 118, S-22100 Lund, Sweden} 

\begin{abstract}
After more than 80 years from the seminal work of Weizs\"acker and the liquid drop model of the atomic nucleus, deviations from experiments of mass models ($\sim$ MeV) are orders of magnitude larger than experimental errors ($\lesssim$ keV). Predicting the mass of atomic nuclei with precision is extremely challenging. This is due to the non--trivial many--body interplay of protons and neutrons in nuclei, and the complex nature of the nuclear strong force. 
Statistical theory of learning will be used to provide bounds to the prediction errors of model trained with a finite data set. These bounds are validated with neural network calculations, and compared with state of the art mass models.
Therefore, it will be argued that the nuclear structure models investigating ground state properties explore a system on the limit of the knowledgeable, as defined by the statistical theory of learning. 
\end{abstract}
%\PACS{21.60.De, 24.10.Ht, 25.40.Dn}
\maketitle  

{\bf Introduction}. Many relevant properties of the atomic nuclei are extremely sensitive to their binding energy, i.e. mass, e.g. decay lifetimes and reaction rates. Therefore, the highest possible precision in reproducing and predicting nuclear masses is needed \cite{Nazarewicz:18}. 
However, current state of the art models deviate from experimental binding energies orders of magnitude more than experimental errors. 
This letter will investigate difficulty of improving the precision of nuclear mass models from an information theory point of view.

The complexity of this problem was hinted in the context of chaotic quantum systems \cite{Aberg:02, Bohigas:02}. That is, the statistical distribution of masses shows a chaotic behaviour is formidable to deterministically reproduce (cf. also \cite{Barea:05}). The amount of nuclear structure data is related to the finite number of nuclear isotopes available in the laboratory and that can possibly exist. In particular, theories that model only the ground state, such as some mass models and density functional theory (DFT), in principle access only few ground state properties of each nucleus.
As of the last atomic mass evaluation \cite{Audi:17}, 3435 nuclei have been measured in the laboratories around the world. Due to the lack of a comprehensive model of nuclear binding energies, it is not known how many nuclei could exist. However, it is safe to assume that this number will not change by an order of magnitude (the current consensus argues that $\sim 7000$ nuclei can possibly exist \cite{Nazarewicz:18}). In practice, many of these exotic nuclei will not be measured in the foreseeable future. 
The effect of the limited number of nuclear masses available can be investigated using statistical learning theory. 

Statistical learning theory deals with the problem of devising a specific predictive model, belonging to a class of models $\Lambda$, using a set of data. When the number of data is finite, there is a limit to the precision that can be reached by a model. Qualitatively, the more a model is \emph{complex} the more data it will need to reach a given predictive power.
This manuscript will analyse the statistical learning bounds of deviation reachable by models attempting to reproduce and predict nuclear binding energies. This is done considering the mass model a statistical learning problem \cite{Vapnik:99} with a complexity constrained by its parametric representation and the precision limited by finite number of available data points. Perceptron networks will be used to validate the statistical learning theory assumptions in this context. Moreover, the bounds for notable density functionals will be examined. The tools here provided enable the analysis of the performance that can be expected from a model when the statistical treatment is rigorous and without bias. This will shed light on the development of functionals and mass models from an information theory perspective.

{\bf Method}. Statistical learning frames within quantifiable boundaries the effect that limited information has in the training of models \cite{Vapnik:95}. The objective of a general learning problem is the minimization of the {\it total risk} functional $R(\alpha)$. That is, finding the set of parameters $\alpha$ by which a given {\it model} best reproduces the data available and predicts the ones that could be taken under consideration. 
However, when working with a finite data set, what is actually minimized is the {\it empirical risk}. That is, the risk evaluated over a finite number of data $l$. Usually the risk is defined using the root mean square deviation (RMSD) of model function or functional $f_\alpha$ respect to the data. $f_\alpha$ takes a set $\alpha$ of parameters belonging to a space $\Lambda$ which defines the class of models under consideration. Therefore, the value of $\alpha$ which minimizes $R_{emp}(\alpha)$ have to be found. 
Under specific conditions, defined by the empirical risk minimization principle (ERM), the minimum of $R_{emp}(\alpha)$ converges (in probability) to the minimum of the total risk functional $R(\alpha)$ when the number of data is large ($l \rightarrow \infty$) \cite{Vapnik:92, Vapnik:99}. Therefore this principle enables a model to make reliable predictions.

However, when the number of data is finite, a good $R_{emp}$ does not guarantee a corresponding predictivity (i.e., good total risk). In the case of limited number data, a model trained on limited data has only a probability of being generalizable for predictions. In this case, the complexity of the model plays a role. A conventional rule of thumb is that, given the same performance on known data, a ``simple'' model generalizes better than a ``complicated'' one. The complexity is often summarily evaluated as the number of free parameters. This is well known as the Occam's razor principle \cite{Blumer:87} (cf. also \cite{Domingos:99}). Statistical learning theory can precisely quantify the impact of the tradeoff between complexity and data availability through the structural risk minimization (SRM) induction principle \cite{Vapnik:95, Taylor:98}.

The degree of complexity of a set of functions can be quantitatively defined using Vapnik and Chervonekis (VC)--dimension \cite{Vapnik:71}. The VC--dimension is, for a set of boolean functions $\Theta_\alpha (x)$ with $\alpha \in \Lambda$, the maximum number $h$ of input vectors $x_1, ..., x_h$ that can be shattered, i.e. separated in the $2^h$ possible ways by the function in set $\Lambda$. The definition can be generalized for a bounded, real model $ a \leq f_\alpha(x) \leq b $, with $a,b \in \mathbb{R}$ (in the mass model case, e.g. $a = 0$ MeV and $b = 8.7945$ MeV is the maximum binding energy per nucleon, that is of $^{62}$Ni isotope), defining a corresponding set of boolean functions,
\begin{equation}
 \Theta_\alpha (x,c) = \theta(f_\alpha(x) - c),
 \label{eq:indicator}
\end{equation}
with $\theta$ the Heaviside unit step function ($\theta(z) = 0$ for $z < 0$, and $\theta(z) = 1$ for $z \geq 0$), and $c \in (0,b)$. The VC--dimension of the set of real valued $f_\alpha (x)$ corresponds to the VC--dimension of the set of the indicator functions $\Theta_\alpha (x,c)$ in Eq. (\ref{eq:indicator}) \cite{Hastie:09}. 
That is, the number of points in $\mathbb{R}$ the related indicator function (\ref{eq:indicator}) can shatter.

% - VC-dimension of polynomial expansion
For example, a lower bound on the VC--dimension of a polynomial in $f: \mathbb{N}^2 \rightarrow \mathbb{R}$ is given by lifting the polynomial to the space of its monomials, and generating a set of point associated with each of the terms of the basis of polynomials. Therefore, is possible to calculate the complexity of this notable polynomial,
\begin{equation}
 E(N,Z) = \sum^N_{i,j = 0} a_{ij} A^i Z^j,
 \label{eq:poly}
\end{equation}
where in mass models $A$ is the total number of nucleons, $Z$ the atomic charge or number of protons, and the binding energy $E$ is parametrized as a polynomial of these variables. The VC--dimension of such polynomial is therefore $h = (N+1)^2$.

% - VC-dimension of neural networks and impossibility to have a decent training through NN
In the following, a sequence of real valued feed--forward neural networks will be used to validate the statistical learning bounds. For such a network, the VC--dimension $h$ was demonstrated to be $O(N) \leq h \leq O(N^2)$ \cite{Maass:95, Koiran:96}, with $N$ the number of weights. 
% In the case of hyperparameter evaluation, the classes of possible functions are explored further increasing the dimension of the hypothesis by at least $N \textrm{ln} N$.

% Given the VC--dimension $h$ of a bounded set of model functions, is possible to bound the difference between the total and empirical risk functions for all the function of the set \cite{Vapnik:99} with probability $1 - \eta$, for ``large'' number of data ($l > 20 h$),
% \begin{equation}
%  R(\alpha) - R_{emp} (\alpha) \leq \frac{b d}{2} \left( 1 + \sqrt{1+ \frac{4 R_{emp}}{b d} }\right),
%  \label{eq:emp_R}
% \end{equation}
% with,
% \begin{equation}
%  d = 4 \frac{h \left(\textrm{ln}(2l) - \textrm{ln}(h) \right) - \textrm{ln}(\eta)}{l}.
% \end{equation}
% Note that for a given $h, \eta > 0$ and a bounded $R(\alpha)$, for $l \rightarrow \infty$ $ R_{emp} (\alpha) \rightarrow R(\alpha)$.
% 

Using VC--dimension to quantify the data complexity of a model, it is possible to derive the minimum amount of data points needed to reach a generalization error with a certain probability. The generalization error is the difference between the RMSD over the training set $R_{emp} (\alpha)$, and the hypothetical RMSD over the whole set of applicability $R (\alpha)$. In other words, if a model class is bounded and has finite VC--dimension, then it is possible to obtain a polynomial bound on the generalization error that this model will have, with a given probability, respect to the number of data provided. This paradigm is known as probably approximately correct (PAC) learning \cite{Valiant:84} and the model is defined as PAC learnable.

The lower bound of data points needed to learn a binary classifier up to a generalization error was demonstrated in \cite{Ehrenfeucht:89}. The understanding of this case has been recently improved, demonstrating the exact bound \cite{Kontorovich:18}. 
The bound of the minimum number number of examples needed to reach a given generalization error was demonstrated also for bounded functions in $\mathbb{Z}$ \cite{Haussler:92}, and it has been extended to include noisy data \cite{Bartlett:96}. 
In this work the bound will be derived from the Hoeffding's inequality \cite{Hoeffding:63}. Therefore, the number of data points $m$ needed to possibly reach a generalization error $\epsilon$ with probability $\delta$ in a PAC learning setting is at least
\begin{equation}
m \geq \frac{1}{\epsilon}\left[ \textrm{ln}(h) + \textrm{ln}\right( \frac{1}{\delta} \left) \right],
\label{eq:m}
\end{equation}
with $h$ the VC--dimension of the model class under consideration. It is of notice that there is no guarantee that the function $f_\alpha$ with $\alpha \in \Lambda$ with error $\epsilon$ exists, but only a $1 - \delta$ probability.
Therefore, in the case of a mass model, there is probability $1-\delta$ that within our hypothesis space $\Lambda$ exists a function where the training error is $\epsilon$ away from the total error over all nuclei in the whole nuclide chart up including the ones not yet discovered.

%%%%%--- Results -----%%%%%

\emph{\bf Results}. Eq. (\ref{eq:m}) can be now used to estimate the number of data needed for a given precision. That is, to estimate the minimum expected error $\epsilon$ using $m$ data points to train a model with VC-dimension $h$. The dataset used consists of the 2016 Atomic Mass Evaluation \cite{Audi:17} (AME16), considering all nuclei with $N, Z \geq 8$, including the phenomenological estimates, for a total of 3336 nuclei and associated masses.

%%--        NN         --%%

At first, to validate the PAC--learning bounds in this context, several feed--forward neural networks with different properties have been trained on some fraction of the AME16 data. The network must take $A,Z$ integer doublets and give back an $\mathbb{R}$ number that represents the binding energy $E$. Therefore, the network is a model $f_\alpha:\mathbb{N}^2 \rightarrow \mathbb{R}$, with $\alpha \in \Lambda$ are the parameters of the network. The network is composed of an input layer with 50 sigmoid nodes, and a single output node to give $E$. In between a number of hidden nodes and layers with rectified linear unit (reLU) activation function. The number of nodes $n$ and layers $L$ is varied to test different VC--dimensional networks. The number of weights for such a network is $51n +n^L$, therefore its VC--dimension is at least $h \geq 51n + n^L$. Of particular interest in this letter is the case of 1 hidden layer of 1000 nodes, denoted in the following as $NN1$, selected as example with $h \geq 52000$. The structure described has been chosen after a hyperparameter optimization for good performance. Rectified linear unit has been chosen for its piecewise linear structure, that guarantees the neural network dimensionality bounds of \cite{Maass:95, Koiran:96}, and for performance aligned with other activation function (cf. supplementary material).

\begin{figure}[h]
 \centering
 \includegraphics[width=0.45\textwidth]{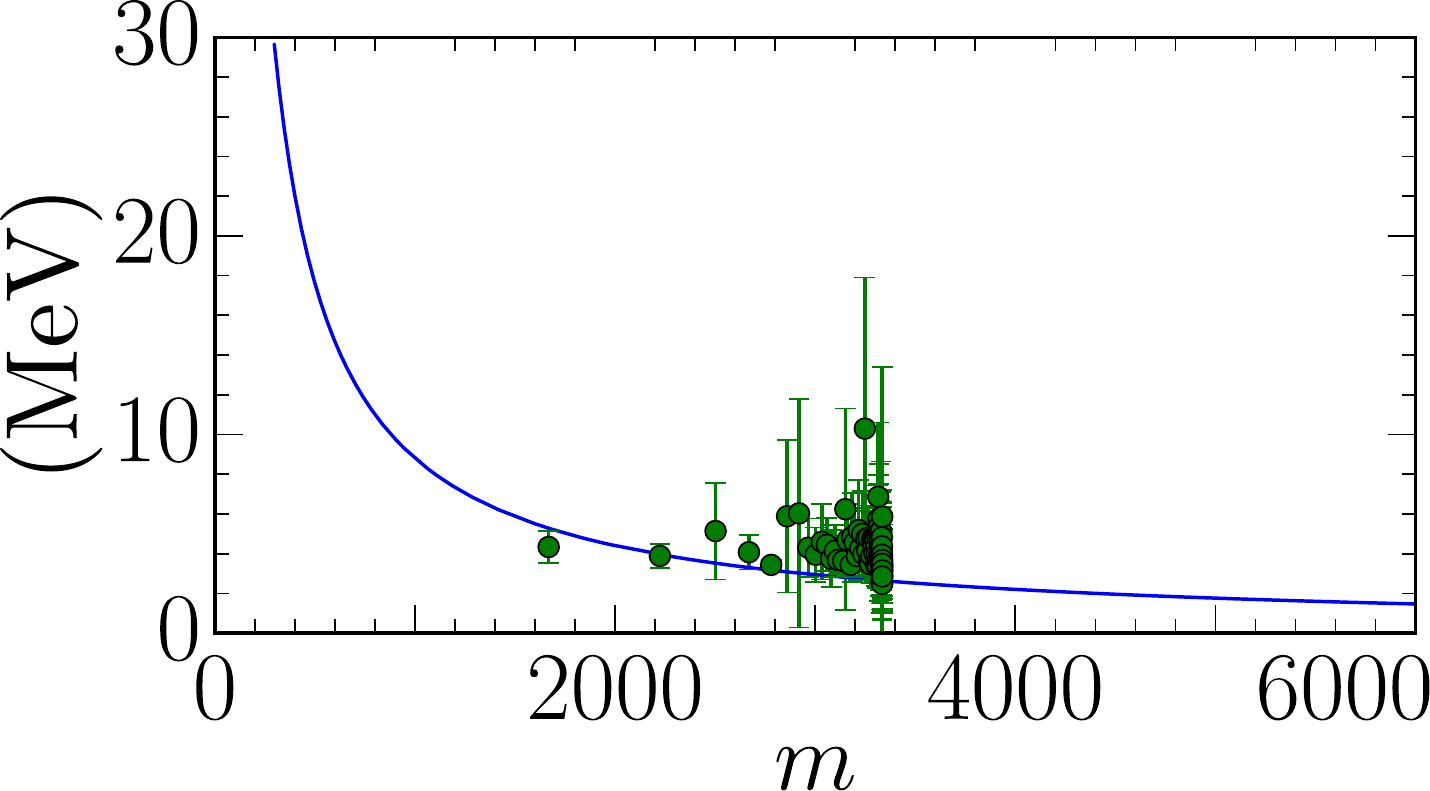}
 \caption{(Color Online) Root mean square deviation of $NN1$ on the $k$--fold cross validation of the AME16 dataset (points). Related PAC--learning lower bound of the generalized error $\epsilon$ from (\ref{eq:m}) (line) in function of the number of data $m$.}
 \label{fig:1}
%git commit a2cfd9213803441caf23a38b5d75f8498d3a3d42
\end{figure}

\begin{figure}[h]
 \centering
 \includegraphics[width=0.45\textwidth]{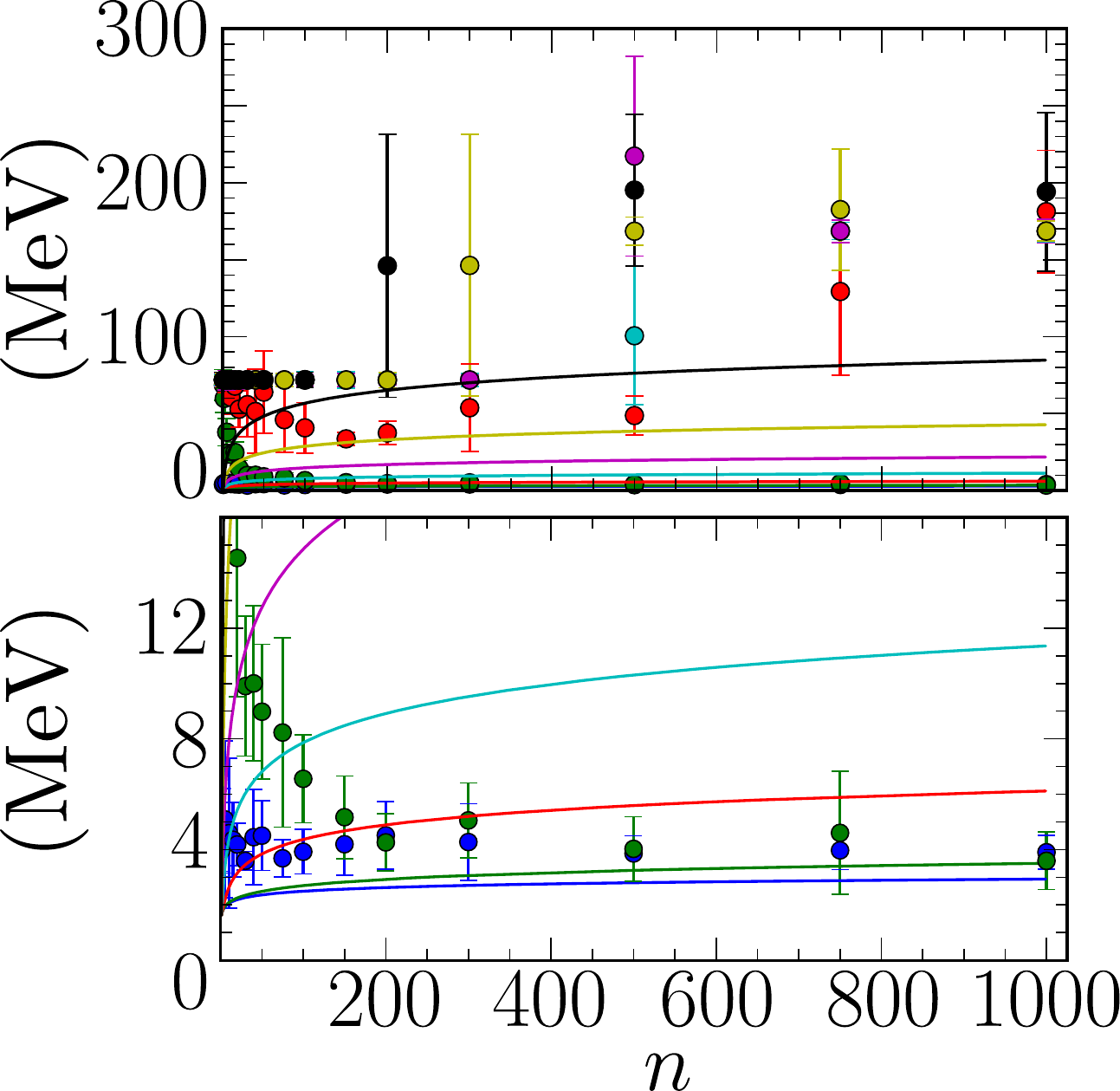}
 \caption{(Color Online) Comparison of error lower bound and results of neural network training with a given number of nodes $n$ and hidden layers, figure (top) and close-up at small error/deviation (bottom). Root mean square deviation of a feedforward neural network on $10$--fold cross--validation of the AME16 dataset (points). The neural network consist of 50 input layer, 1 output layer, a number of hidden layers specified by the following color coding: blue (1), green (2), red (4), cyan (8), magenta (16), yellow (32), black (64). Each of the hidden layers has $n$ nodes.
 $\epsilon$ corresponding lower bounds from (\ref{eq:m}) with VC--dimension defined as directly proportional to number of weights (line).}
 \label{fig:2}
%git commit a2cfd9213803441caf23a38b5d75f8498d3a3d42
\end{figure}

The training loss is evaluated and optimized as RMSD between calculated and experimental binding energies in a training set. The RMSD in Figs. \ref{fig:1}, \ref{fig:2} are evaluated on the validation set, using $k$-- fold cross validation technique \cite{lachenbruch:68}. This guarantees an assessment of the fitting procedure which reduces the bias in the test/validation selection \cite{efron:83}. The uncertainty related to the RMSD is calculated as standard deviation of different folds. The results presented in Figs. \ref{fig:1}, \ref{fig:2} testify to the reliability of PAC learning bounds for this system, the RMSD of models for different neuron number approaches the PAC limit of $\epsilon$ when the network can be trained to reliably describe the system. This result is useful to relate the VC--complexity of a mass model, and the training data provided, with its error. In the following we will use this result to relate VC--complexity of nuclear models with their expected performance.

%%--   Semi-Empirical   --%%

The Weizs\"acker semi--empirical mass formula \cite{Weizsacker:35,Bethe:36} is one of the first attempts to describe the binding energy of an atomic nucleus in function of powers of the number of nucleons $A$ and protons $Z$ (cf. supplementary material). It is of notice that its fractional contributions are not directly related to Eq. (\ref{eq:poly}). However, it is straightforward to relate this function to a VC--dimension of at least $((N=6) + 1)^2$, if we consider with no bias all the possible combinations of radius, surface, volume and symmetry terms, or 6 if we consider the parametrization derived with the physical bias.

%%--        DFT        --%%

Density functional theory is used to describe systems composed of many quantum particles \cite{Sholl:09}, and has been an extremely successful model of atomic nuclei \cite{Bender:03,Ring:80}. It is based on the Hohenberg-Kohn theorems \cite{Hohenberg:64}, 
%that is: i) ground state properties of a many--fermion system are uniquely determined by the fermion density, and ii) there exists one and only one density that minimize the functional, that is the ground state density and its energy density functional returns the ground state energy. This latter preposition is known 
and the {\it variational principle}. That is, the model functional of density $\rho$, $E[\rho(x)]$, will be minimized varying the densities with some Lagrangian constrains (e.g. that the densities contain the correct number of particles) and its minimum will correspond to the exact ground state density and corresponding energy.

The functional $E_\alpha[\rho(x)]$ is usually a complicated combination of densities, eventually derived from an pseudo--potential \cite{Perlinska:04,Bennaceur:17}. The parameters $\alpha$ of the pseudo--potential are tuned to reproduce physical ground state properties. The densities and properties are calculated through the variational principle. Therefore, the same principle of risk minimization and consequent bounds applies. However, to calculate the exact VC--dimension of a complicated functional is not trivial due to the non--linearity of the operation the functional applies on the density. But a conservative lower bound can be provided considering that, 
\begin{equation}
  h(E[\rho(x)]) \leq h(E(x)),       
  \label{eq:h}
\end{equation}
this will allow us to calculate a lower bound on the VC--dimension of popular functionals, that in turn determines a lower PAC bound to the generalization error and number of data.

The very popular Skyrme density functional is composed of a contact interaction, with a momentum--dependent term (which translates in derivatives of the densities in the functional) and a density dependent term. The functional can be related to a polynomial expansion (plus the density dependent term) of the density using (\ref{eq:h}), reducing to the VC--complexity of a second order polynomial \footnote{first order derivatives go to 1, and other terms cancel} over two dimensions (neutrons and protons) and 8 constrains on the parameters, therefore with dimension at least $h_{\textrm{Skyrme}} \geq 2((N=2)+1)^2 - 8 = 10$. 
As a title of example, in the case of the Gogny functional \cite{Decharge:80} the pseudo--potential is composed by two Gaussians with different widths and a density--dependent term. The Gaussian itself has VC--dimension 3, and there are 8 terms for each, making the VC--dimension at least $h_{\textrm{Gogny}} \geq 24$.

Interestingly, theorem 6.8 and following of \cite{Vapnik:95} state that a good rate of convergence can be reached only for smooth functions. Despite being derived from pseudo--potentials with difficult discontinuities (e.g. Skyrme pseudo-potential is a combination of Dirac $\delta$) the resulting densities are smooth and therefore can be converged. 
% But how stable this convergence might be in light of the above theorem might be studied in the future.

\begin{table}[h]
\begin{center}
\begin{tabular}{l c c c c}
model            & VC--dim & RMSD [MeV]               & $\epsilon$ [MeV] & $\tilde m$ \\ 
\hline
Weiszacker       &    6    & $3.41 \pm 0.19$              &  1.09   &  36434 \\ 
                 &    49   &                              &  1.45   &  48395 \\ 
$NN1$            & $52000$ & $4.22 \pm 1.06$              &  2.64   &  88076 \\ 
Skyrme (UNEDF0)  &    10   & 1.428 \cite{Kortelainen:14}  &  1.18   &  39343 \\ 
Gogny (D1M)      &    14   & 0.798 \cite{Goriely:09}      &  1.24   &  41259 \\ 
\hline\hline
\end{tabular}
\caption{Properties of different models to describe nuclear physics masses. The columns represent i) the lower bound on VC--dimension for the given model; ii) RMSD of the referred models; iii) lower bound on the $\epsilon$ error provided by the Hoeffing inequality in PAC--learning considering 3336 homogeneously weighted data points; iv) number of data points $\tilde m$ needed to reach a generalization error $\epsilon$ of 100 keV with 99\% probability, a considerable improvement to current bounds. To be noted that the RMSD result for Gogny D1M in \cite{Goriely:09}, contains beyond DFT corrections. RMSD for $NN1$ is obtained averaging all the results in Fig. 1.
}
\label{table:1}
\end{center}
\end{table}

%%% Points to make %%%

A long standing problem in the creation of nuclear density functionals and mass models is the number and type of data that has to be included in their fitting. This is particularly important in the study of next generation, high order, nuclear density functionals \cite{Bennaceur:17,Davesne:13}. This work moves towards quantifying the amount of data needed to reach an expected precision in a given mass model. 
Table \ref{table:1} shows a comparison of known mass models RMSD and PAC--learning bounds. The interpretation of Table \ref{table:1} comes with several \emph{caveats}. 

The RMSD related to Skyrme UNEDF0 and Gogny D1M are calculated on the data available in the atomic mass evaluation 2003 \cite{Audi:03}. Moreover, in the case of Skyrme UNEDF0 only on even--even nuclei. By definition a RMSD on a limited amount of data is the empirical risk and cannot be considered a generalization error, even more so when calculated on data belonging to the training set. The generalization error $\epsilon$ in Table \ref{table:1} is derived from the PAC learning bounds considering training on 3336 homogeneously weighted data points, corresponding to the available measured binding energies in AME16 \cite{Audi:17}. 
Most importantly, the objective of such functionals such as UNEDF0 and Gogny D1M is not only to reproduce masses but other properties as well, optimizing a complicated cost function. Physical bias enters in the construction of this cost function, even when the statistical approach is rigorous such as in the UNEDF program \cite{Kortelainen:10} and to a greater extent in the definition of other successful models.
Sometimes functionals have been developed as interactions, considering a leaving room for beyond mean field correlations in the total energy and other observables in beyond mean field approaches. The RSMD value cited for Gogny D1M especially, includes beyond DFT physics and additional corrections. 
To be noted also that in some models parameters can be redundant or not sensitive to the observables under consideration. E.g. in the Weiszacker semi--empirical mass formula surface and Coulomb is easy to see parameters are highly correlated \cite{Pastore:19}, and this is also true for DFT models \cite{Kortelainen:10, Haverinen:17}. This, in principle, decreases the VC dimensionality with respect to the observable under consideration, but also the post-- and pre--dictive power of the model.
Despite these \emph{caveats}, the close values of the bound $\epsilon$ and the RMSD of state of the art models, testifies to the possibility here discussed of investigating the limits of precision in mass models through statistical learning as defined by PAC learning. Moreover, considering the number of data needed for a significant improvement $\tilde m \sim 4 \times 10^4$ according to PAC learning bounds, the relation between $\epsilon$ and RMSD appear robust to the addition of few more data points.

\emph{\bf Conclusions}. This work is just a first step in the evaluation of statistical learning precision bounds in many--body systems, and further investigations on different models and more complete properties will be required to analyze different many--body models. However, some conclusions can be drawn with better hindsight than before possible.
In light of this work on statistical learning theory, the difficulty in further improving mass models to reach a predictive and precise estimate of nuclear masses might not be a shortcoming of some specific model. Instead, the necessary information for a model with predictivity is unlearnable on the basis of masses (or few ground state properties) alone. 
To increase the performance of these models, especially in biasless, statistically robust next generation functionals, a variety of observables must be included. Eventually, investigating the response to fields will be crucial. This will involve ab-initio calculations (that is, interaction including nucleon-nucleon scattering data) or unification of structure and reactions.

% On the other hand, data available has an intrinsic bias towards stable nuclei and nuclei with more binding energy, making the approach to unbound nuclei more challenging. The boundaries between bound and unbound nuclei are so--called drip lines, and are experimentally out of reach for the most part. In fact, the position of the drip lines (that can be regarded as a classification boundary) would be one of the most important prediction of a mass model. 
% In some models parameters can be redundant, e.g. in the Weiszacker semi--empirical mass formula surface and Coulomb parameters are highly correlated \cite{Pastore:19}. This decreases the VC dimensionality but also the post-- and pre--dictive power of the model.

This work for the first time has investigated the VC--dimension related to a many-body method and its implication regarding the performance that a given model and related training can reach. This work suggests that many--body methods might not only be judged by their computational complexity, as in the novel field of Hamiltonian complexity \cite{Whitfield:13, Gharibian:15}, but also in terms of their information complexity represented by VC--dimension and PAC learning bounds.

% - Some points on DFT and Hamiltonians
% - The development of DFT/HF functional is a Merlin-Arthur problem.
% - Conclusions : - Hopeleness of polynomial expansion, neural network fitting or other bias-less methods. Especially through Hyperparameter fitting
%                 - Suggestion that DFT is also a difficult way to solve the problem
%                 - Need of Excitations and possibly reaction data
% - Outlook : VC--dimension of many-body methods
%%%%%%%%%%%%%%%%%%%%%
\subsection{Acknowledgment}
The Quadro P6000 GPU used for this research was donated by the NVIDIA Corporation. This work benefited from discussions with the participants to the workshop ``Novel approaches for the description of heavy nuclei'' organized at Lund University 19-21 March 2019, with the contribution of Newton Alumni Fellowship of the Royal Society.
The source code used will be included in the supplementary material of the publication and publicly released on a git server at a later date.

\section{Supplementary}

\subsection{Methods}
The adopted geometry of the neural network and training: 50 input nodes, 1 output nodes and training making use of 0.5 dropout of the hidden layers have been obtained as the most consistent values after a tree Parzen estimate hyperparameter optimization (based on Expected Improvement method). Other than sigmoid activation function, also rectified linear unit and softmax have been tried in all the possible combination between input, hidden layers and output. The resulting RMSD was not significantly impacted, testifying to the robustness of the PAC--learnable boundary. Rectified linear unit in the hidden layer has been chosen for simplicity in calculating the VC dimensionality.

The root mean square deviation (RMSD) has been calculated in {\it cross--validation} \cite{lachenbruch:68}. It consist in dividing the dataset into training and validation exclusive subsets. The procedure is repeated several times with different separation of training and validation dataset guaranteeing a bias-free assessment of the fitting procedure which improves on Bootstrap method \cite{efron:83}. This method consist in:

\begin{itemize}
 \item Divide the training set in a number of equivalent subsets $k$ (usually, but not necessarily, randomly picked). This makes up the $k$--fold. A popular option, empirically verified to perform well in a variety of situation, is $k=10$. 
 \item Train the set on a set composed of $k-1$ subsets, and validate it on the remaining one.
 \item Repeat the training $k$ times, so that training and validation are considered over all the possible validation sets.
 \item From the RMSD resulting from the combination of training--validation, consider the average and the standard deviation of RMS deviations.
\end{itemize}

The average RMSD and its deviation will inform on the performance of the model and cost function chosen, and its resilience to modification of the dataset and therefore predictive power. Where otherwise not specified, $k=10$ has been used. Other $k$--fold choices were also tested, including a ``complete cross-validation'', that is a number of folds equal to the number of data.

The Weiszacker model represented in Table 1 of the main article, has been obtained with a RMSD optimization over AME12 database \cite{Audi:12} and validated using the AME16 database \cite{Audi:17}. The cost function adopted is the modified $\chi^2$,
\begin{equation}
 \tilde{\chi}^2 = \sum_i ( f_{\alpha}(x_i) - y_i )^2/ (\textrm{log}(\Delta y_i))^2,
  \label{eq:chi}
\end{equation}
that has been adjusted in the case measured data $y_i$ have errors $\Delta y_i$ that span different orders of magnitude, as for the case of errors in mass measurements, from \cite{Dobaczewski:14}.

\subsection{Mass Model}
\begin{figure}[h]
 \centering
 \includegraphics[width=0.45\textwidth]{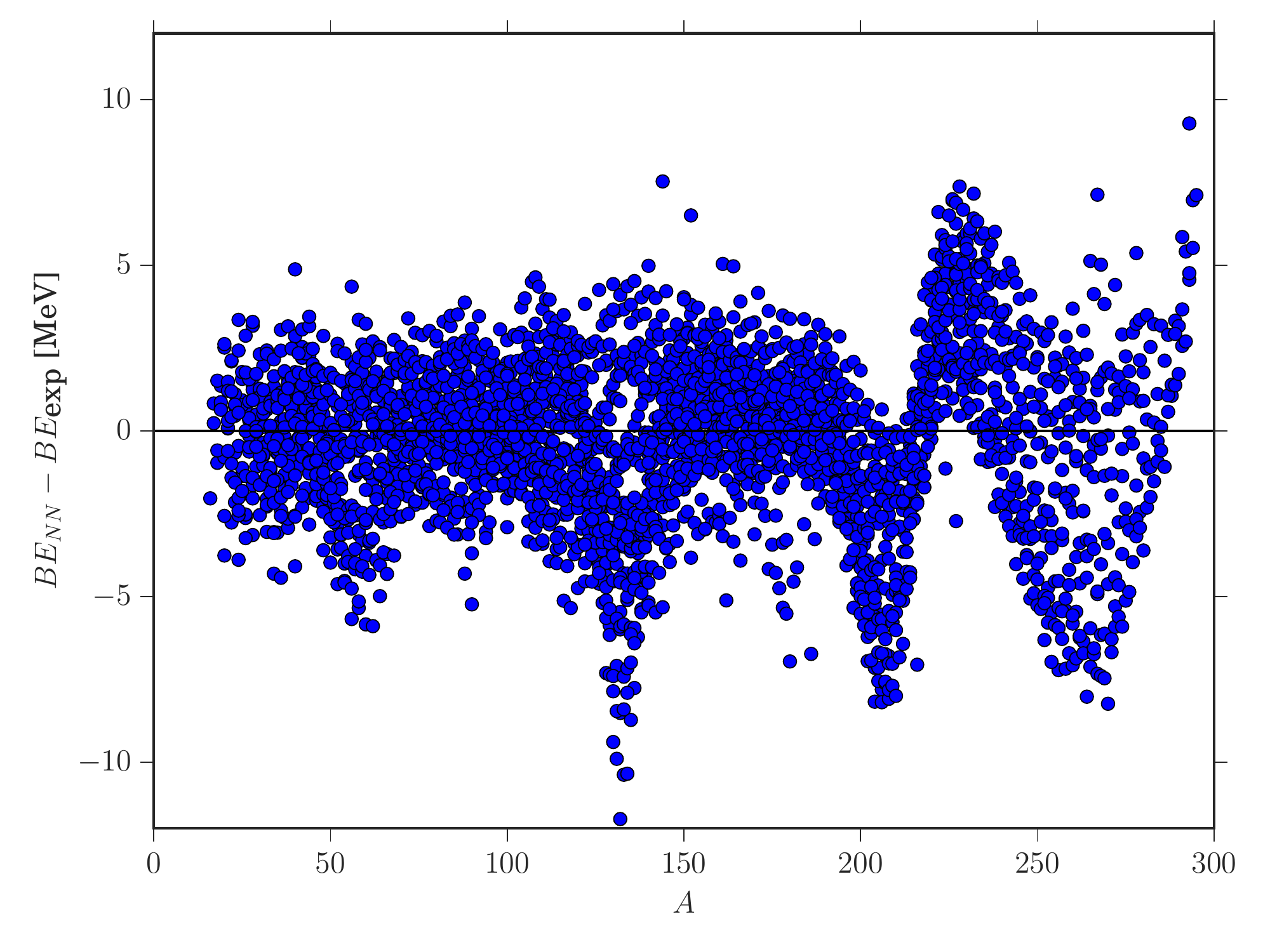}
 \caption{(Color Online) Difference between the neural network postdiction and experimental binding energy in MeV in function of the atomic mass number. The neural network is composed of 1000 rectified linear unit nodes in the hidden layer, is the best resulting out of the 10--fold cross validation over AME12 dataset. Similarly to most mass models, the familiar arches in correspondence of the magic numbers are present.}
 \label{fig:3}
%git commit a2cfd9213803441caf23a38b5d75f8498d3a3d42
\end{figure}

\begin{figure}[h]
 \centering
 \includegraphics[width=0.45\textwidth]{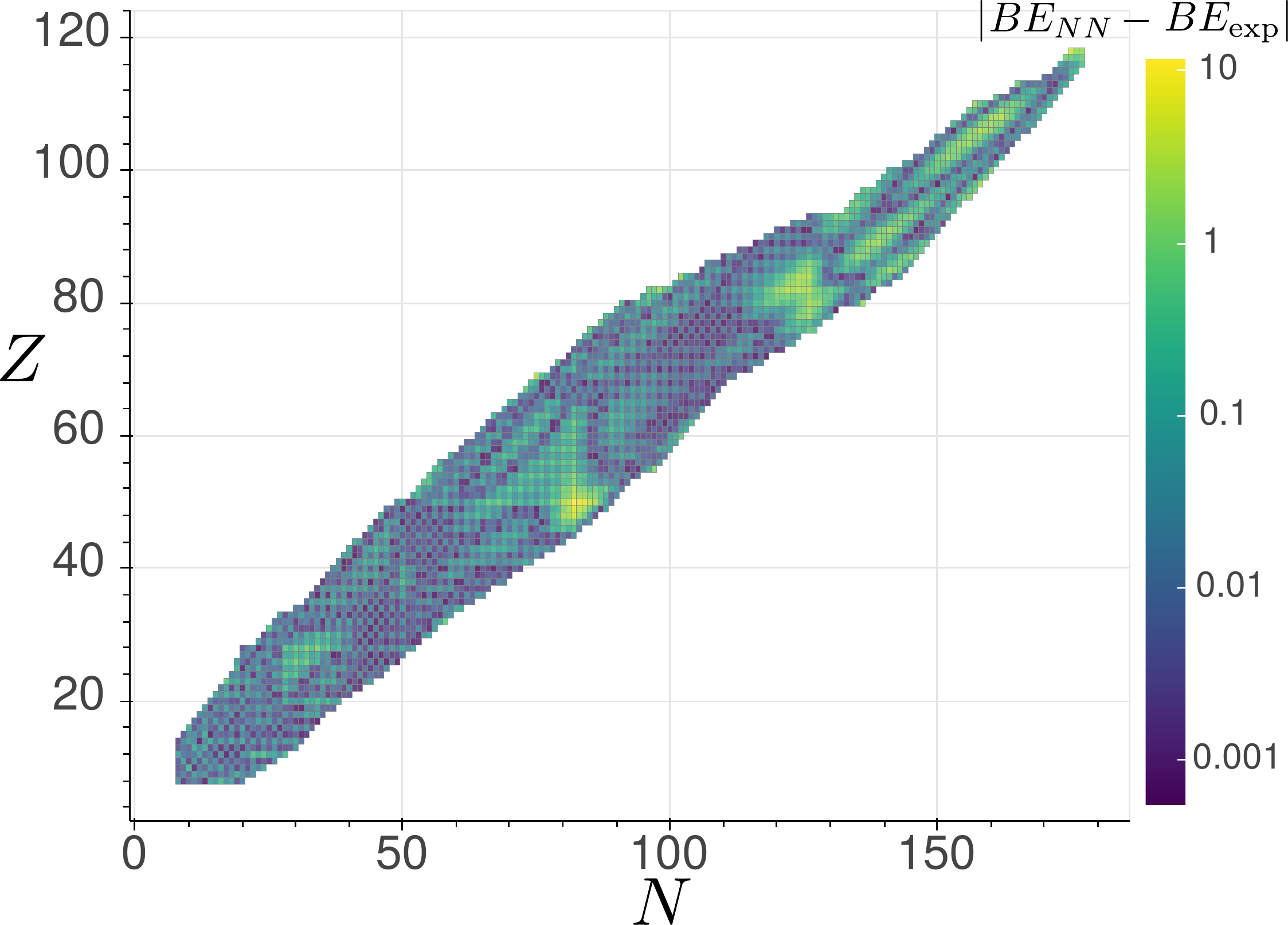}
 \caption{(Color Online) Segre chart of the isotopes with the difference between the neural network postdiction and experimental result in MeV. The neural network is composed of 1000 rectified linear unit nodes in the hidden layer, is the best resulting out of the 10--fold cross validation over AME12 dataset. That results in a root mean square deviation on the AME16 dataset of 2.54 MeV.}
 \label{fig:4}
%git commit a2cfd9213803441caf23a38b5d75f8498d3a3d42
\end{figure}

% A. 2 and 5 MeV
% G. 3.6 and 0.725 MeV
Neural network approaches to fit to nuclear masses have been tried with specific configurations for neural network structure and propagation strategy \cite{Gernoth:93, Athanassopoulos:04}, obtaining results between 0.7 and 5 MeV of RMSD on masses on specific tests set. The result of this work, obtained with a randomly initialized model that does not introduce bias through the geometry, is 2.54 MeV on the larger AME16 dataset in Figs. \ref{fig:3} and \ref{fig:4}. To be noted that the best results in \cite{Gernoth:93} are obtained with the networks with the least number of weights, i.e. lower VC--dimension.

Also models to reproduce HFB calculations \cite{Bayram:14} and other nuclear properties, such has radii \cite{Akkoyun:13}, have been recently introduced. Moreover, neural networks have recently emerged as possible method for providing additional corrections and correlations (arguably representing missing shell--effects) to a previously devised mass--model with excellent results \cite{Utama:17,Niu:18}.

The Weiszacker semi--empirical mass formula has been studied for comparison,
\begin{align}
  E (A,Z) = & a_v A - a_s A^{2/3} - a_C \frac{Z(Z-1)}{A^{1/3}} - a_a \frac{(A-2Z)^2}{A} \notag \\
                  & + a_p \frac{\delta_p}{\sqrt{A}},
  \label{eq:mass}
\end{align}
where $\delta_p$ is 1 for $A$ and $Z$ even, 0 for $A$ or $Z$ odd, -1 for $A$ and $Z$ odd. The coefficients of the of the semi--empirical model mass formula, consist in the well known volume $a_v$, surface $a_s$, Coulomb $a_C$, asymmetry $a_a$ and pairing $a_p$ terms. The parameter regarding the results used in Table 1 of the main article are reported here in Table \ref{table:2}, with a striking resemblance in quantity and uncertainties with \cite{Pastore:19} (results obtained independently and with different fitting procedures). The related uncertainties have been computed with the covariance matrix calculated as inverse of the Hessian, which is composed by derivatives of the cost function respect to the free parameters. By linear approximation it is possible to estimate errors related to a parameter or observable \cite{Toivanen:08,Dobaczewski:14, Bennaceur:17}

\begin{table}[h]
\begin{center}
\begin{tabular}{l}
$a_v   =  15.40 \pm 0.014$ \\
$a_s   =  16.71 \pm 0.042$ \\
$a_C   =  0.701 \pm 0.001$ \\
$a_a   =  22.56 \pm 0.037$ \\
$a_p   =  11.88 \pm 0.823$ 
\end{tabular}
\caption{Coefficients in MeV of the semi--empirical mass formula (\ref{eq:mass}) obtained fitting AME12. Notice the uncertainties related to the different quantities, especially concerning pairing that testifies the softness of the regression of Weizsacker mass formula respect to pairing, that is also the reason why different formulations are used for this term. }
\label{table:2}
\end{center}
\end{table}

Support vector machine \cite{Li:05} obtained interesting results of accuracy and would be also an interesting object of study in light of PAC learning bounds which are well established \cite{Vapnik:95}.

\bibliographystyle{apsrev4-1}

\bibliography{../../nuclear.bib,../../neural.bib}

\end{document}